# Successful strategies for competing networks

J. Aguirre[1*], D. Papo[2] and J.M. Buldú[2,3*]

[1]Centro de Astrobiología, CSIC-INTA, ctra. de Ajalvir km 4, 28850 Torrejón de Ardoz, Madrid, Spain, [2]Center for Biomedical Technology (UPM), Campus de Montegancedo, 28223, Pozuelo de Alarcon, Madrid, Spain, [3]Complex Systems Group, URJC, C/Tulipán s/n, 28923, Móstoles, Spain, *email: aguirreaj@cab.inta-csic.es; javier.buldu@urjc.es

**Competitive interactions represent one of the driving forces behind evolution and natural selection in biological and sociological systems[1,2]. For example, animals in an ecosystem may vie for food or mates; in a market economy, firms may compete over the same group of customers; sensory stimuli may compete for limited neural resources in order to enter the focus of attention. Here, we derive rules based on the spectral properties of the network governing the competitive interactions between groups of agents organized in networks. In the scenario studied here the winner of the competition, and the time needed to prevail, essentially depend on the way a given network connects to its competitors and on its internal structure. Our results allow assessing the extent to which real networks optimize the outcome of their interaction, but also provide strategies through which competing networks can improve on their situation. The proposed approach is applicable to a wide range of systems that can be modeled as networks[3].**

Often, the outcome of a competitive process between agents is affected not only by their direct competitors but also by the specific network of connections in which they operate. Complex networks theory offers a large number of topological measures[3], which can be derived in a simple way from the adjacency matrix *G*, containing the information on network connectivity. These measures can then be used to explain important dynamical and functional properties such as robustness[4,5], synchronization[6], spreading[7] or congestion[8,9].

Hitherto, the emphasis has been on the properties of single isolated networks. Typically, however, networks interact with other networks, while simultaneously retaining their original identity. Recently, a few studies examined how global structural properties or dynamical processes taking place on them are affected by the existence of connected networks, showing for instance that robustness[10-13], synchronization[14], cooperation[15,16], transport[17] or epidemic spreading[18-21] change dramatically when considering a *network of networks*[22]. Interestingly, while certain types of network interdependencies may enhance vulnerability with respect to the case of isolated networks[10], the addition of links between networks may also hinder cascading processes, such as failures in a power grid, on interconnected networks[23]. This holistic view requires a broad redefinition of classical network parameters[24]. Even more importantly, one of the major challenges lies in the identification of guidelines for *how to best link networks*[25].

If one considers the outcome of interaction from the perspective of a single network, which is either forced into or evaluates the benefits from interacting with another one, an important challenge arises: "How can I make the most of my interaction with another network?" To answer this question, we consider two separate networks competing for given limited resources, which in general are related to the structural properties of the network and to the outcome of a certain dynamical process. The two networks interact by creating common links, but retain their original identity. Inasmuch as the competition process can be thought of as a struggle for "importance", one way to quantify its outcome is in terms of *centrality*, a network measure of importance. Here, we propose to use a particular measure of centrality: *eigenvector centrality* (see Supplementary Section S1.1). Eigenvector centrality quantifies the importance of a node's neighbours, and can be high either because a node has numerous or important neighbours[26]. In an ensemble of networks, the centrality of each network is simply the sum of the centralities of all its nodes.

Eigenvector centrality is more than just a measure of topological importance since it is directly related to dynamical processes occurring on networks. Suppose that the state $n_i(t)$ at time $t$ of each node $i$ in a network undergoes a dynamical process, and that the vector state $n(t)$ describing the whole network evolves according to $\vec{n}(t+1) = M\vec{n}(t)$, where $M$ is a process-specific transition matrix with $M_{ij} \geq 0$. Under rather general conditions (see Supplementary Section S1.2 and ref. 27), $n(t)$ evolves towards an asymptotic state described by the first eigenvector $\vec{u}_1$ of $M$, and the asymptotic growth rate is given by its highest eigenvalue, $\lambda_1$. Interestingly, eigenvector centrality is also calculated as the first eigenvector $\vec{u}_1$ of the transition matrix $M$. In addition, when all link weights are equal, it coincides with the first eigenvector of the adjacency matrix[26] $G$. Therefore, studying eigenvector centrality is tantamount to studying the steady state of dynamical processes on networks.

We study the simplest case of two competing undirected networks (see Supplementary Section S3 for a generalization to directed networks, where links have a unique direction, or to $N$ coupled networks). Suppose that two networks ($A$ and $B$) respectively of $N_A$ and $N_B$ nodes and $L_A$ and $L_B$ links, are connected via a set of $L$ connector links, to create a new network $T$ of $N_T=N_A+N_B$ nodes and $L_T=L_A+L_B+L$ links (see Fig. 1). The centrality $C_A$ of network $A$ is given by the sum of the components of the first eigenvector $\vec{u}_{T,1}$ of $M_T$ corresponding to the nodes belonging to network $A$. This quantity can be normalized by the sum of the centrality of all nodes $C_A = \sum_{k=1}^{N_A} (\vec{u}_{T,1})_k \big/ \sum_{k=1}^{N_T} (\vec{u}_{T,1})_k$, which yields $C_T=1=C_A+C_B$, so that networks $A$ and $B$ simply vie to increase their share of centrality in the overall system $T$.

Our first objective is to find an expression of $\vec{u}_{T,1}$ that depends explicitly on the parameters of networks $A$ and $B$ before connection and on the set of tentative connector links. This way, we can have *a priori* knowledge of the final distribution of centralities and prescribe an adequate strategy of connection. $\vec{u}_{T,1}$ can be expressed as the sum of $\vec{u}_{A,1}$, the eigenvector centrality of network $A$ prior the connection, and a term resulting from the connection of network $A$ with network $B$. In other words, network $T$ can be thought of as a perturbation of graph $A$ by graph $B$, with connector links having a weight $\varepsilon \ll 1$. Disregarding second order terms, $\vec{u}_{T,1}$ can be approximated as (see Supplementary Section S2):

$$\vec{u}_{T,1} = \vec{u}_{A,1} + \varepsilon \sum_{k=1}^{N_B} a_k \vec{u}_{B,k} + o(\varepsilon^2) \qquad [1]$$

$$a_k = (\vec{u}_{A,1} P \vec{u}_{B,k}) \big/ (\lambda_{A,1} - \lambda_{B,k}) \qquad [2]$$

where $\vec{u}_{A,1}$ and $\lambda_{A,1}$ are respectively the first eigenvector and corresponding eigenvalue of matrix $M_A$ of network $A$, and $\vec{u}_{B,k}$ and $\lambda_{B,k}$ are the $k$ eigenvectors and eigenvalues of matrix $M_B$ of network $B$ (and, as is customary, for generic cases $\varepsilon$ could be replaced by unity[28]). These equations are valid when network $A$ is the strong one, that is, when it has the highest first eigenvalue associated with it ($\lambda_{A,1} > \lambda_{B,1}$). The connector links are included in a perturbation matrix $P$, with $P_{ij} \neq 0$ for links connecting both networks and $P_{ij}=0$ otherwise (where $i$ and $j$ represent the node indices of networks $A$ and $B$ respectively). Because the first term of Eq. 1 is fixed, the outcome of the competition only relies on the second term, which is dominated by (a) the product $\vec{u}_{A,1} P \vec{u}_{B,1}$, which is proportional to the eigenvector centralities of the connector nodes (i.e. the nodes reached by connector links) before connection, and (b) the highest eigenvalues associated to both networks, $\lambda_{A,1}$ and $\lambda_{B,1}$. We start by focusing on how the choice of the connector nodes influences the outcome of the competition. According to their importance in the network, certain nodes can be classified as peripheral (low centrality) and central (high centrality). Therefore, if we focus on the importance of connector nodes, we can consider four different *connecting strategies* when creating links between networks: peripheral-to-peripheral (*PP*), peripheral-to-central (*PC*), central-to-central *(CC)* and central-to-peripheral (*CP*) (see Fig. 1A). Keeping in mind that the strong network (that is $\lambda_{A,1} > \lambda_{B,1}$), equation (1) allows extraction of the general rules, or successful strategies (see Supplementary Section S2.3), showing network A and B how to optimize their centrality when interacting with each other.

Firstly, connecting the most peripheral nodes of networks $A$ and $B$ minimizes the term $\vec{u}_{A,1} P \vec{u}_{B,1}$ and therefore optimizes the centrality of network $A$. On the contrary, connecting the most central nodes maximizes $\vec{u}_{A,1} P \vec{u}_{B,1}$ and optimizes the centrality of the competitor $B$. Secondly, increasing the number of connector links reinforces the weak competitor $B$, as more positive terms are added in the perturbation matrix $P$, leading to an increase of $\vec{u}_{A,1} P \vec{u}_{B,1}$. In fact, by varying the connector links, network centrality is dramatically modified by several orders of magnitude (see Fig. 1B and Supplementary Fig. S1 for details).

Turning to the second quantity affecting competition, Eqs. 1 and 2 show that, to prevail in generic cases, a network must increase its associated $\lambda_1$. One way to achieve this is to increase its number of nodes and/or internal links, a fact illustrated in Fig. 2A and thoroughly studied in Sections S4 and S5. We stress that the goal of each competitor is not to overcome its opponent, but to optimize its centrality gain resulting from interaction, as unless $\lambda_{A,1} \approx \lambda_{B,1}$, from the spectral considerations studied herein the network with higher eigenvalue always retains an absolute advantage. Importantly, the strategy followed to choose the connector nodes determines the amount of benefit (Fig. 2A), a phenomenon that is particularly noticeable for large networks with close values of $\lambda_1$ (see inset of Fig. 2A).

Often, networks are constrained in the way they connect to each other or are simply reluctant to modify their connector nodes. In addition, increasing network size may not be practicable. Think for example of interacting spatial networks with boundary constraints and, in particular, of the transoceanic air-navigation networks, whose connector hubs are given airports. Under these conditions, the only way to increase $\lambda_1$ is to reorganize the internal structure. An increase in $\lambda_1$ can be achieved by appropriately rewiring the network internal links. To overcome the strong network $A$, the weak network $B$ needs not necessarily rewire in the optimal way proposed by Ref. 29; all it needs to do to prevail in the competition is to attain a situation where $\lambda_{B,1} > \lambda_{A,1}$ (see Fig.

2B). Clearly, the nature of the connector nodes affects the rate at which the transfer of centrality from one network to the other occurs when the weak network becomes the strong one: peripheral connections lead to a drastic transition when $\lambda_{B,1}$ overtakes $\lambda_{A,1}$, favoring a winner-takes-all situation; on the contrary, connecting central nodes leads to smoother but continuous changes. Furthermore, these results are general as the topological structure of the network only impacts the smoothness of the transition to the winning scenario (see Fig. 2C).

Suppose that we now want to evaluate how well a real network has played its cards in the competition process. Take for instance the well-studied dolphin community of Doubtful Sound[30] (see Fig. 3 and Supplementary Section S6). In this example, sixty-two bottlenose dolphins are grouped into two networks interacting through some *connector individuals*. Unavoidably, the question of which network is taking advantage from the actual choice of connectors arises. This can be quantified by reshuffling the connector links and evaluating the competition parameter $\Omega$:

$$\Omega = \frac{2\left(C_A - C_A^{\min}\right)}{C_A^{\max} - C_A^{\min}} - 1, \qquad [3]$$

where $C_A^{max}$ ($C_A^{min}$) is the maximum (minimum) centrality that network $A$ can achieve through the rewiring of connector links. $\Omega=0$ when none of the networks benefited from the real distribution of connector links, whereas $\Omega=1$ (-1) indicates that the strong (weak) network took full advantage of the competition. Note that $\Omega$ indicates whether a network benefits from the actual distribution of connector links, not which network has higher centrality.

Figure 3A shows that in the observed situation the strong network $A$ competed in a rather advantageous way ($\Omega=+0.70$). On the other hand, the weak network $B$ could have improved on the actual situation: though never sufficient to make it prevail in absolute terms, a different wiring would have increased network $B$'s centrality by a factor of six (Fig. 3B).

Finally, in real world situations, the effects produced by connecting networks are not instantaneous. The rules sketched above can be thought of as steady-state ones. One way to quantify the time the system takes to feel the effects of the connections is by considering that the dynamics in the whole network, comprising $A$ and $B$ and their interconnections, evolves according to $\vec{n}(t+1) = M\vec{n}(t)$. Under these conditions, the competition time $t_{c,T}$ to reach steady state is given by the function:

$$t_{c,T} \propto \left(\ln\frac{\lambda_{A,1}}{\lambda_{B,1}}\right)^{-1}\left[1+\varepsilon^2 f\left(\lambda_{A,k},\lambda_{B,k},\vec{u}_{A,k},\vec{u}_{B,k},P\right)\right] \qquad . \qquad [4]$$

The first term shows that $t_{c,T}$ is maximum for $\lambda_{B,1} \approx \lambda_{A,1}$. The last term, which in practice is dominated by $\lambda_{A,1}, \lambda_{B,1}$ and the product $\vec{u}_{A,1} P \vec{u}_{B,1}$, accounts for the differential effects of the particular type of connection between networks. Figure 4 shows that competition times are always longer for peripheral-to-peripheral than for central-to-central connections, an effect that grows dramatically when approaching $\lambda_{B,1} = \lambda_{A,1}$ (see Supplementary Sections S2 and S4 for a full analytical treatment of this phenomenon).

The general rules through which centrality is redistributed as a result of competitive interactions allow *describing* competition among networks with a high degree of generality. Because centrality is, in turn, related to a number of dynamical processes taking place on networks, the range of possible applications includes phenomena as diverse as population distribution (see Supplementary Sections S1.2, S1.3 and S3.2), epidemics (see Supplementary Section S1.4), rumor spreading (see Supplementary Section S1.5), or Internet navigation (see Supplementary Section S3.1). For example, in the case of the evolution of populations of RNA molecules, one could quantify the competition for genotypes showing a given phenotype, and cast light on how evolvability and adaptability are related to the as yet poorly understood genotype-phenotype correspondence (see Supplementary Section S3.2). Importantly, our results also allow *prescribing* successful competition strategies, and engineering desirable connectivity patterns. For instance, the administrator of a website could design an adequate hyperlinking policy to attract the maximum number of visits from competing domains.

## Acknowledgments

The authors acknowledge the assistance of D. Peralta-Salas in the analytical computations, fruitful conversations with A. Arenas, S. Boccaletti, C. Briones, J.A. Capitán, F. del-Pozo, J. García-Ojalvo, J. Iranzo, C. Lugo, S.C. Manrubia, M. Moreno, Y. Moreno, G. Munoz-Caro and A. Pons, and the support of the Spanish Ministerio de Ciencia e Innovación and Ministerio de Economía y Competitividad under projects FIS2009-07072, FIS2011-27569 and MOSAICO, and of Comunidad de Madrid (Spain) under Project MODELICO-CM S2009ESP-1691.

## Author contributions:

J.A. and J.M.B. had the idea behind the new methodology proposed in the paper. J.A. and J.M.B. developed the theory and performed the numerical simulations. J.A., J.M.B. and D.P. participated in the motivation and discussion of the results. J.A., J.M.B. and D.P. wrote the manuscript.

## Additional Information:

The authors declare no competing financial interests. Supplementary Information accompanies this paper (please obtain it free of charge from
http://www.nature.com/nphys/journal/vaop/ncurrent/extref/nphys2556-s1.pdf ).

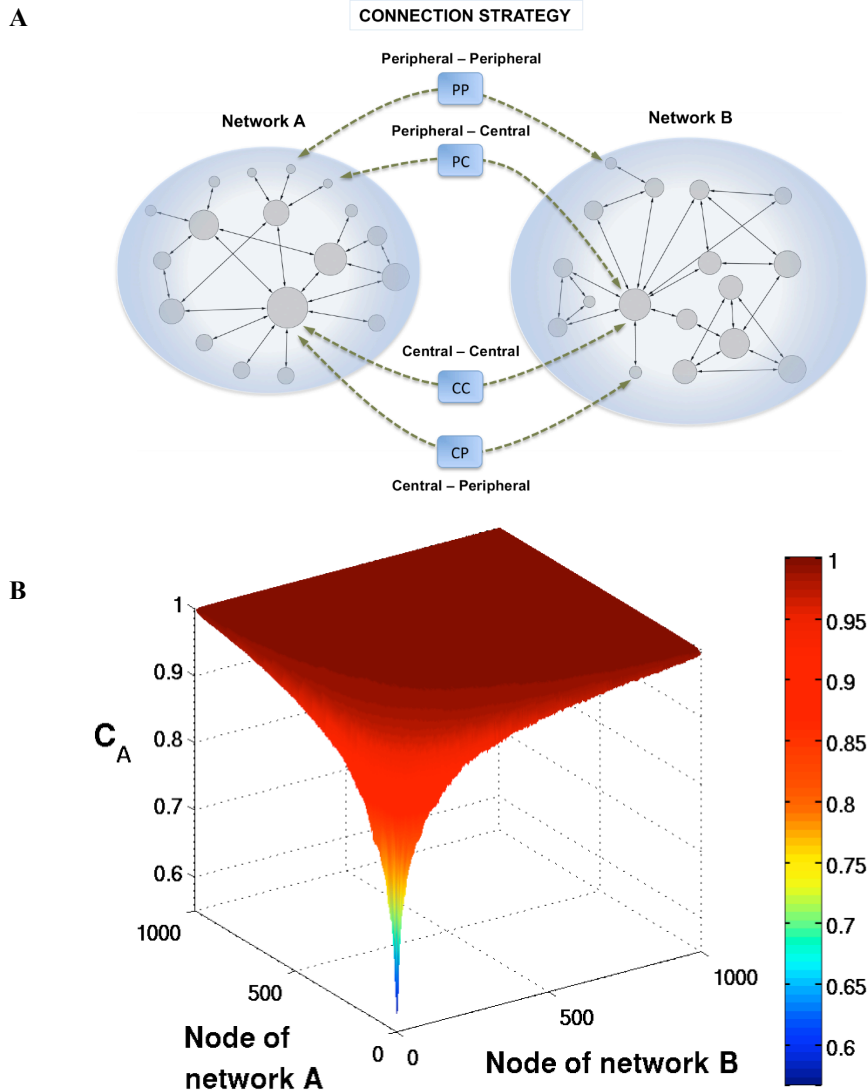

**Fig. 1. Connection strategies and outcomes: representative diagram and numerical example of competition for centrality between networks.** A) Schematic representation of the different strategies for creating connection paths between two undirected networks. Central nodes (*C*) and peripheral nodes (*P*) are respectively those with higher and lower eigenvector centrality, a measure of importance. Initially, networks remain disconnected before one or more connector links are added. The role played by the connector nodes, i.e. the nodes reached by connector links, determines four possible strategies. In this example, nodes' sizes are proportional to eigenvector centrality before connection. B) Competition between two Barabási-Albert networks[3] *A* and *B* of size $N_A=N_B=1000$ and $L_A=L_B=2000$ links, connected by one single connector link in all possible configurations ($10^3 \times 10^3$). $C_A$ is the centrality of network *A* ($C_B=1-C_A$). Note that $C_B$ varies from $10^{-5}$ to 0.44 depending on the connector link. The axes represent the connector nodes in networks *A* and *B*, and nodes are numbered according to their network centrality rank. In this example, $\lambda_{A,1}>\lambda_{B,1}$, indicating that *A* is the strong network. Thus, connecting central nodes benefits the weak network *B*.

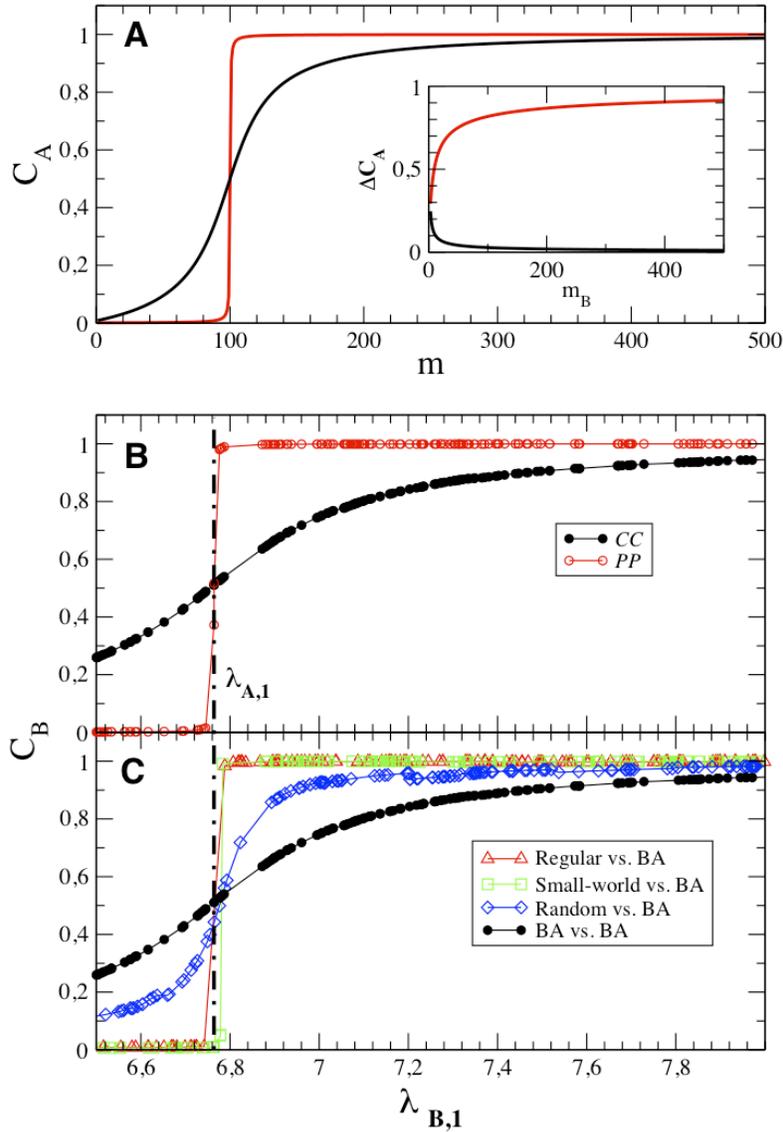

**Fig. 2**. **Competition strategies based on increasing the maximum eigenvalue associated to the competing networks.** A) A star network $A$ of $m$ nodes competes, increasing its size, against a star network $B$ of $m_B=100$ nodes. The network centrality $C_A$ depends on the size $m$ and on the strategy used to create the connections between both networks (black lines correspond to central-central $CC$ and red lines to peripheral-peripheral $PP$). Inset of A) shows the increase of $C_A$ ($\Delta C_A$) from $m=m_B-1$ to $m=m_B+1$ as a function of the network size $m_B$. Note that $\Delta C_A$ is directly related to the slope of $C_A$ in $m=m_B$ and, therefore, it measures the rate of the transition of the centrality from one competitor to the other. B) Competition between two Barabási-Albert (BA) networks[3] of the same size ($N_A=N_B=200$ nodes and $L_A=L_B=400$ links), where network $B$ reorganizes its internal topology to increase $\lambda_{B,1}$ and overcomes network $A$ ($\lambda_{A,1}=6.764$). Again, the connecting strategies are crucial in the outcome. C) Similar results are obtained when networks reorganize departing from different initial structures. In this case, only $CC$ strategy is shown. In all three competitions A), B) and C) only one connector link was used.

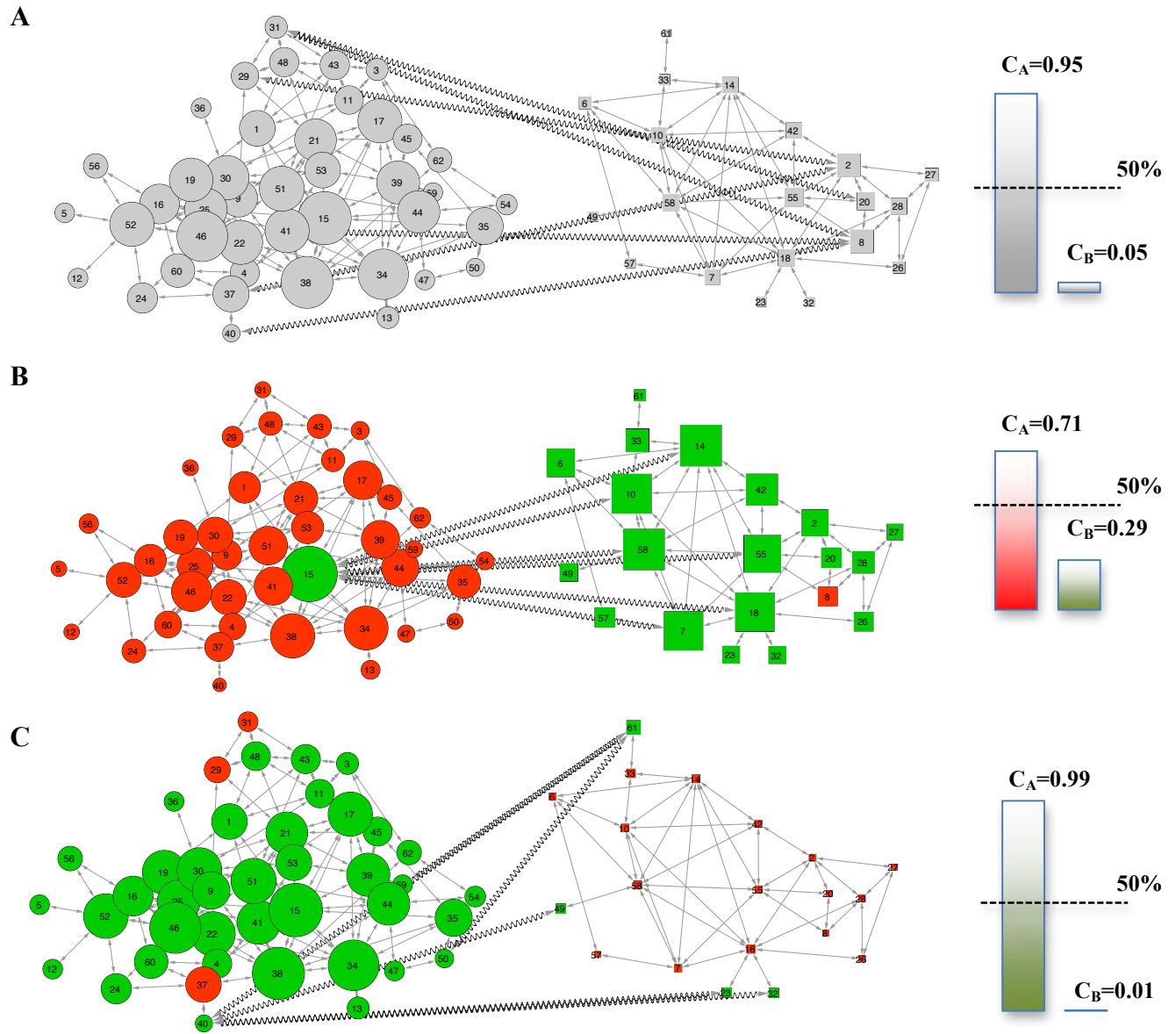

**Fig. 3. Competition in real networks: the case of the dolphin community of Doubtful Sound[30].** A) Real distribution of centrality in the dolphin network. The shape of the nodes shows the community they belong to: circles for network *A* and squares for network *B*. Connector links between networks are indicated by sine-shaped links. The size of the nodes is proportional to their centrality. B) Distribution of centrality where connector links are distributed in a way that maximizes centrality of network *B*. Colors indicate if nodes are increasing (green) or decreasing (red) their centrality with regard to the real case. Note that it is possible that some nodes do not have the same positive/negative outcomes as the networks they belong to. C) Same procedure as in B) but, in this case, connector links maximize centrality of network *A*. The increase of centrality of network *A* introduced by the connecting strategy ($\Omega = +0.70$) may not always be desirable. For example, if a disease spreading process were considered, network *A* would be affected faster (see Supplementary Section S1.4 for details).

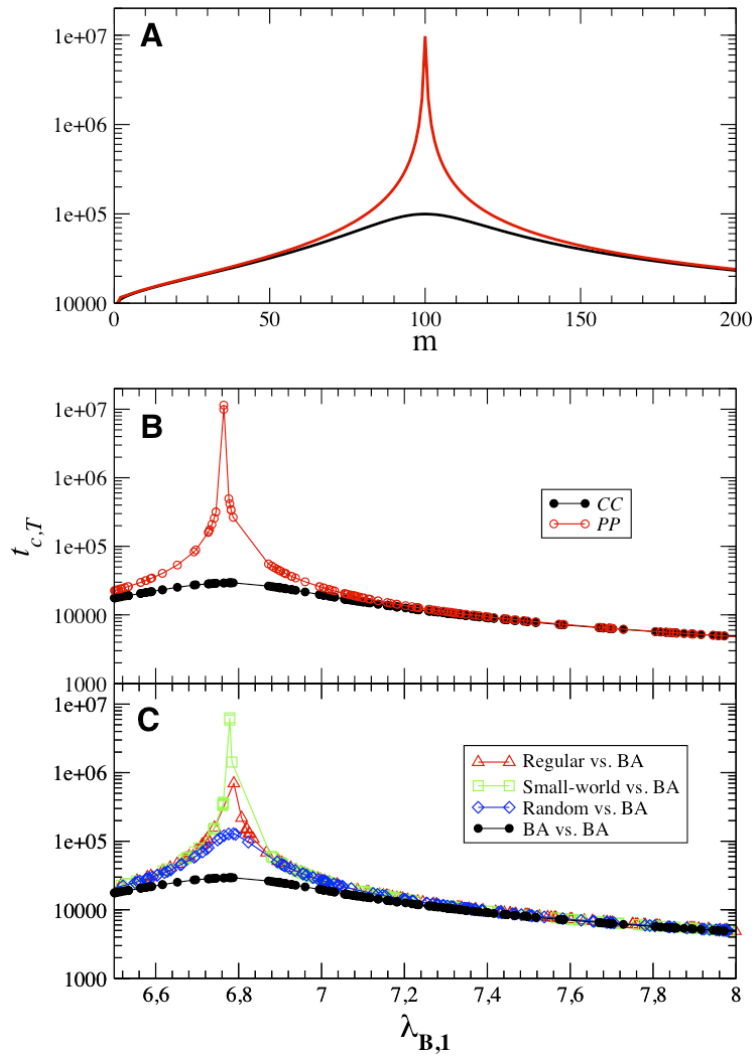

**Fig. 4. Competition time $t_{c,T}$ to reach the steady state in a competition process between two connected networks** (see caption of Fig. 2). A) A star network $A$ of $m$ nodes competes, increasing its size, against a star network $B$ of $m_B=100$ nodes. Competition time $t_{c,T}$ depends on the size $m$ and on the strategy used to create the connections between both networks (black lines correspond to central-central $CC$ and red lines to peripheral-peripheral $PP$). Connections between peripheral nodes always lead to much longer competition times, a fact that must be taken into account in the connecting strategy. B) Competition between two Barabási-Albert (BA) networks[3], with $\lambda_{A,1}=6.764$ and network $B$ underlying an internal rewiring process. Again, the connecting strategies are crucial in the time to reach the steady state. C) In this case, network $B$ reorganizes departing from different initial structures, affecting the competition time. In all three competitions A), B) and C) only one connector link was used.